# Protein Microarrays with Carbon Nanotubes as Multi-Color Raman Labels

Zhuo Chen,[1,4] Scott M. Tabakman,[1,4] Andrew P. Goodwin,[1] Michael G. Kattah,[2] Dan Daranciang,[1] Xinran Wang,[1] Guangyu Zhang,[1] Xiaolin Li,[1] Zhuang Liu,[1] Paul J. Utz,[2] Kaili Jiang,[3] Shoushan Fan,[3] & Hongjie Dai[1]

**Detection of biomolecules is important in proteomics and clinical diagnosis and treatment of diseases. Here, we apply functionalized, macromolecular, single-walled carbon nanotubes (SWNTs) as multi-color Raman labels to protein arrays for highly sensitive, multiplexed protein detection. Raman detection utilizes the sharp peaks of SWNTs with minimal background interference, affording a high signal-to-noise ratio needed for ultra-sensitive detection. Surface-enhanced Raman scattering (SERS) combined with the strong resonance Raman intensity of SWNTs, affords detection sensitivity down to 1 fM, a three order of magnitude improvement over most of reported fluorescence-based protein detections. We show that human autoantibodies to Proteinase 3 (aPR3), a biomarker for the autoimmune disease Wegener's granulomatosis, is detected by Raman in human serum up to a 1:10[7]**

[1]Department of Chemistry and Laboratory for Advanced Materials, Stanford University, Stanford, CA 94305, USA. [2]School of Medicine, Stanford University, Stanford, CA 94305, USA. [3]Department of Physics, Tsinghua-Foxconn Nanotechnology Research Center, Tsinghua University, Beijing 100084, China. [4]These authors contributed equally to the work. Correspondence should be addressed to H. D. (hdai@stanford.edu).

**dilution. Moreover, SWNT Raman tags are stable against photobleaching and quenching, and by conjugating different antibodies to pure $^{12}$C and $^{13}$C SWNT isotopes, we demonstrate two-color SWNT Raman-based protein detection in a multiplexed fashion.**

Various methods have been developed for protein detection, including enzyme-linked immunosorbent assays, fluorescence-based protein microarrays,[1] electrochemistry,[2] label-free optical methods,[3,4] surface-enhanced Raman scattering (SERS),[5,6,7] microcantilevers,[8] quantum dots,[9,10] and nanotube[11,12] or nanowire[13] based field-effect transistors. Among them, protein microarrays[14,15] are unique, providing high throughput, multiplexed protein detection, and thus have found wide application ranging from proteomics to disease research and diagnosis.[14,16,17] The typical sensitivity limit of protein arrays based on fluorophore tags is ~1 pM, limited by background interference due to species on the substrate or autofluorescence of reagents.[18] Increasing the sensitivity of protein detection in arrayed format could enhance the capability of this novel method for proteomics research. Moreover, improved understanding of soluble biomarkers in diseases such as cancer has yielded clinical demands for high sensitivity and selectivity, thus allowing for minimally invasive risk assessment, early stage disease diagnosis, and monitoring of responses to therapeutic interventions.[19]

A few diverse methodologies are under development which show promise for highly sensitive protein detection in research and clinical applications. Label-free nanowire-based transistors[13] have demonstrated femtomolar sensitivity, but the sensitivity is limited to samples in pure water or low salt solutions, and not in serum. Another methodology, amplified detection based upon multi-functional nanoparticles, has demonstrated even higher sensitivity,[10] however this method's complexity requires a multitude of reagents and is highly time consuming. The methodology described herein, based upon Raman scattering, is relatively simple, may be easily multiplexed, and exhibits high sensitivity in clinically relevant samples over a large dynamic range, from ng/mL to fg/mL.

The SERS effect provides the potential for rapid, high throughput, sensitive protein detection. Previously, SERS was applied for immobilized protein detection[5,6] by coupling small Raman-active dyes to gold nanoparticles functionalized by ligands. However, the utility of these sensors was limited in sensitivity as typical Raman labels have weak intensities, owing to small Raman scattering cross-sections.[20] As a result, sensitivity is not quantitative[5] or is limited to nM range[5] which does not compare favorably with fluorescence methods. For high sensitivity Raman sensing with dye molecules,[7] long acquisition times are needed and molecules are easily subject to degradation under laser radiation.

Single-walled carbon nanotubes (SWNTs), on the other hand, are ideal labels for SERS-based protein detection. SWNTs have unique one-dimensional structure, and exhibit distinct electrical and spectroscopic properties, including strong and simple resonance Raman signatures. SWNTs posses enormous Raman scattering cross-sections ($\sim 10^{-21}$ cm$^2$ sr$^{-1}$ molecule$^{-1}$), simple, tunable spectra, and are more stable than other organic Raman labels.[21,22] Various schemes have been explored to develop SWNTs for chemical, biological and medical applications.[23-27] Carbon nanotubes have been utilized as in vivo[28] and in vitro[29] optical probes for biological imaging, yet their potential as Raman-tags for highly sensitive detection applications has not been explored previously.

We describe the synthesis and preparation of biocompatible, highly selective SWNT-IgG conjugates for in vitro protein detection. Novel methodology for obtaining uniform surface-enhanced Raman scattering over large areas has been developed and combined with well known self-assembly chemistry for robust protein immobilization coupled with high sensitivity Raman scattering detection.

Herein, we report the use of SWNTs as macromolecular Raman labels for highly-sensitive and selective arrayed protein detection, with 1000-fold greater sensitivity than fluorescence, over 7 to 8 decades of dynamic range for systems demonstrating high affinity interactions. We demonstrate detection of the clinically-relevant human autoimmune disease biomarker anti-Proteinase 3 (aPR3) in human serum down to $1:10^7$ dilution. Multiple-color SWNT Raman labels are developed using isotopic SWNTs, and employed for sensing in a multiplexed fashion requiring only a single excitation source.

**RESULTS**

**Assay design for protein detection by SWNT Raman-tags**

A secondary antibody, goat anti-mouse immunoglobulin-G (GaM-IgG), was conjugated to highly water soluble, short (~50-150 nm, see **Supplementary Fig. S1** online), macromolecular SWNTs, functionalized with PEGylated phospholipids[26] (PL-PEG, see Methods). GaM-IgG conjugation imparted binding specificity of SWNT-tags to mouse antibodies (**Fig. 1**). Protein immobilization on substrates was performed in arrayed fashion, by either covalent attachment to Au-coated glass or on commercial microarray slides by standard robotic spotting (see Method). We developed a novel 6-arm-branched carboxylate-terminated PEG, grafted onto Au-coated surfaces for protein immobilization (**Fig. 1a**), that afforded excellent protein attachment and high resistance to non-specific binding (NSB) effects. Proteins, such as polyclonal mouse IgG or human serum albumin (HSA), were immobilized on the assay surface and either detected by Raman scattering upon binding of GaM-IgG-conjugated SWNTs (using 785 nm excitation laser, see Methods), or were used in sandwich assays, to capture analyte protein (e.g., anti-HSA IgG raised in mouse) from dilute serum (**Fig. 1b**).

To enhance Raman scattering intensity following direct or sandwich assay detection of analyte by SWNT labeled GaM-IgG, we annealed the gold-coated substrate in a reducing hydrogen atmosphere to aggregate the Au film into particles (**Supplementary Fig. S2** online for atomic force microscopy (AFM) images). Alternatively, a 5 nm layer of pure silver was deposited (**Supplementary Fig. S3** online) onto the assay surface at ambient temperature (see Methods). Both techniques led to reproducible SWNT Raman signal enhancement of about 60-fold without damaging SWNTs, as evidenced by the observed strong, characteristic SWNT radial breathing mode (RBM, $< 500$ cm$^{-1}$), longitudinal G$^+$ and transverse G$^-$ mode (near 1590 cm$^{-1}$) signals (**Fig. 1c**) uniformly over the substrate. The strong SERS effect was attributed to field enhancement by surface plasmons of metal particles, formed uniformly over the entire substrate, in near-resonance with the 785 nm laser used for Raman scattering measurements. (**Supplementary Fig. S4** online).[30] For quantitative protein detection, the SWNT G-mode scattering intensity was utilized, as it demonstrates a high signal-to-noise ratio and very narrow peak width ($\sim$20 cm$^{-1}$ FWHM).[31]

**Selective and sensitive detection of model proteins by SWNT-Raman tags**

The selectivity of a protein assay depends on both effective antibody conjugation and the minimization of NSB. In this study, single-walled carbon nanotubes were functionalized by a 1:1 mixture of phospholipid-branched-methoxyPEG (DSPE-3PEO), which provided excellent NSB resistance[11,27,32] (see **Supplementary Fig. S5** online), and linear phospholipid-PEG-NH$_2$ (DSPE-PEG-NH$_2$, see Methods), which provided sites for bioconjugation. To evaluate the selectivity of this assay, different proteins were immobilized on PEGylated Au/glass substrates and tested for binding with GaM-IgG/SWNTs through Raman detection (**Fig. 2a**). We observed specific GaM-IgG/SWNT binding only to the six mouse IgG's immobilized on the substrate, and not to the other eight negative control protein spots (**Fig. 2b**). The application of SWNT Raman tags in biomolecule detection is highly generalizable to any system with adequate binding affinity and specificity (A summary of biomolecules employed in SWNT Raman sensing and their affinities is provided in **Supplementary Table S1** online). In addition to Raman tags for protein assays, SWNTs have been utilized as tags for the in vitro Raman labeling of various cell membrane receptors,[22,26,27,29,32] including both low and high affinity interactions, such as those between cyclic RGD peptide with the cell adhesion molecule $\alpha_v\beta_3$ integrin and anti-CD20 (Rituxan®) with CD20 respectively (see **Supplementary Fig. S6, Fig. S7, Fig. S8, and Fig. S9** online for examples of specific detection over a wide range of binding affinities, including SWNT Raman tags used in biomolecular detection based upon streptavidin-biotin interaction, complimentary ssDNA binding, Protein A and Protein A/G – goat IgG interaction, and in vitro cyclic RGD peptide labeling of cell surfaces).

To explore the sensitivity limit of SWNT Raman-based protein detection, human serum albumin (HSA) and GaM-IgG/SWNTs were used as model capture and reporting agents for sandwich assay detection of monoclonal mouse anti-HSA IgG (aHSA) spiked into fetal bovine serum (FBS, **Fig. 1**). **Figure 3a** shows Raman mapping images of HSA spots after exposure to various concentrations of aHSA from 100 pM to 1 fM followed by incubation with GaM-IgG/SWNTs. The images were generated by plotting the integrated SWNT G-band intensity at each point (50 μm x 50 μm pixel size) over one quarter of the protein spot. In the maps, uniform SWNT signals were observed within HSA spots exposed to aHSA at concentrations above ~1 pM. At lower concentrations, the SWNT signal was sparse, consistent with a small number of aHSA IgGs captured by the HSA layer. Defining the limit of detection (LOD) as twice the standard deviation above the control (without analyte), we reproducibly obtained aHSA detection sensitivity down to 1 fM, over 8 orders of dynamic range (**Fig. 3b**, replicate sets of data are shown, obtained on independent assay chips with different batches of GaM-IgG/SWNT conjugates). The data exhibited sigmoidal or S-shape behavior,[33] suggesting saturation and steric hindrance effects for detection at high concentrations, and increased proportional influence of residual NSB effects at the lower detection limit.

To compare our SWNT-Raman-based detection to standard fluorescence-based protein microarray methods, we performed both assays in parallel. Protein microarray experiments were carried out on Super-Epoxide II slides (Array-It) by replacing the GaM-IgG/SWNT tags with cyanine-3 (cy3) labeled GaM-IgG. A LOD of ~1 pM of aHSA was reproducibly obtained with arrayed fluorescence detection (**Fig 3c,d and Fig. S10**), and this sensitivity limit was comparable to previous observations.[1] For fluorescence-based protein microarrays, we found that the fluorescence signals from HSA spots exposed to ≤ 100 fM analyte were similar to those measured at protein spots without any exposure to Cy3 labeled GaM-IgG. This indicates that for detection of aHSA IgG concentrations ≤ 100 fM, fluorescence intensity was comparable to the background noise of substrates and reagent molecules, thus limiting the sensitivity to ~1 pM. Conversely, in control experiments we observed no proteins or substrates exhibiting Raman peaks at the SWNT G-band position (**Supplementary Fig. S11** online), providing little background interference. Reduced background contribution and improved signal-to-noise owing to bright Raman scattering spectra and surface-enhancement techniques provides an extended dynamic range in comparison to fluorescence-based techniques. Thus, the sharp, surface-enhanced, and background-free Raman scattering peaks observed in our detection method afforded ~1000-fold improved sensitivity over fluorescence methods in this model protein system.

It should be noted that the inflection point slope of a dose-response curve is expected to be unity. This is nearly observed for the fluorescence-based assay of aHSA (slope of 0.88), while SWNT-Raman tags show an inflection point slope of 0.60. This deviation from theory may be due to non-linear effects of surface enhanced Raman scattering[34]. Despite such effects, the observation of sufficiently large Raman signal change down to 1 fM is significant, and can be used to measure low concentrations. Curve fitting is a standard technique we employed to quantify our data in this concentration region, and sigmoidal (4-parameter logistic) regressions have adequately fit the data and allowed quantification.

To exemplify this, a blind study was conducted in which three samples with unknown concentrations of aHSA were prepared and run in parallel with standards. Each data point value was determined by averaging 9-duplicate microarray protein spots on the same assay substrate (**Fig. 4**). The average intensity was taken as the average of the individual mean scattering intensities (n=9), and error bars were calculated from the standard deviation between the mean scattering intensities of the nine duplicate spots. Such treatment yielded a calibration curve well fit by a logistic regression (**Fig. 4**) over 6 orders of magnitude along with mean scattering intensities for the three unknowns and an analyte-free control. Note that only six decades of analyte concentration were used for the calibration curve to avoid overcrowding of the substrate, and possible cross-contamination. The three unknowns were prepared as 5 pM, 200 fM and 5 fM analyte solutions. Sensitivity down to the fM range, within a factor of two to three was observed, with the three blind unknown analyte samples quantified experimentally as 2 pM, 110 fM, and 4 fM respectively, thus demonstrating the accuracy of SWNT Raman-tag detection despite non-linear effects.

**Highly sensitive autoimmune biomarker detection in human serum**

We then applied our SWNT Raman labels to clinically-relevant detection related to Wegener's granulomatosis (WG), an autoimmune disorder associated with anti-neutrophil cytoplasmic antibodies (cANCA).[35] The disease is characterized by multi-organ vasculitis and can be fatal in severe cases. Currently, the gold standard for WG diagnosis and treatment is immunohistological staining; however, the positive predictive value is only ~50%.[35] Autoantibodies directed against Proteinase 3 (PR3), a 29 kDa serine protease, are directly implicated in the pathogenesis of WG, and are used for diagnosing WG. Immunoassays for aPR3 have proven more effective in prediction of WG flares than staining alone.[36] We synthesized goat anti-human (GaH) secondary antibody, GaH-IgG/SWNT, Raman tags for detection of human aPR3 in human serum, mimicking the serum of a WG-positive patient. A human IgG sample containing aPR3 isolated from a c-ANCA-positive patient (Immunovision) was diluted 1:100 to 1:10,000,000, all in 1% normal human serum (**Fig. 5a**). A normal human serum sample, without spiked aPR3 (control), was included as a negative control. Detection of aPR3 by the SWNT Raman method in comparison with the negative control sample was successful over more than 7 decades of dilution (**Fig 5b**). A sigmoidal dependence, modelled by four parameter logistic fit, blue curve, was observed, and the inflection point slope was approximately 0.50. Using Cy3 fluorescently-labeled GaH-IgG, the LOD was reached at only 4 orders of dilution of the original aPR3-spiked human serum solution. Beyond that, the signals were comparable to the background noise, measured from the normal human serum control (**Fig. 5c**). A sigmoidal dependence was fit by four-parameter logistic function, red curve (inflection point slope was approximately 0.90).

**Multi-color detection by SWNT Raman-labels**

Finally, SWNT Raman tags may be applied for multi-color detection. This was accomplished by synthesizing $^{12}$C and $^{13}$C isotopic SWNTs by chemical vapor deposition[37,38] using $^{12}$C- and $^{13}$C-pure methane respectively. The Raman mode frequency scales with atomic mass as[39] $\omega \propto m^{-1/2}$. The G-band Stokes' shift of pure $^{13}$C SWNTs ($\omega_{C13}$) relates to that of pure $^{12}$C SWNTs ($\omega_{C12}$=1590 cm$^{-1}$) by $\omega_{C13}$=[1-

$(12/13)^{1/2}]\omega_{C12}$=1528 cm$^{-1}$, as observed experimentally (**Fig. 6b**). By comparing the Raman scattering intensity of $^{12}$C and $^{13}$C SWNT G-bands at their respective maxima, approximately 0.5% cross-talk was observed for the isotopomers (see **Supplementary Fig. S12** online), Through differential conjugation of minimally cross-reactive secondary IgGs, SWNTs could be utilized for detecting two types of IgGs simultaneously. GaM and GaH IgGs were conjugated to $^{12}$C SWNTs ($^{12}$C-GaM) and $^{13}$C SWNTs ($^{13}$C-GaH), respectively. A mixture of the two conjugates was incubated on the sensing assay surface, leading to differential binding to mouse and human primary IgGs with high selectivity (**Fig. 6a, Supplementary Fig. S13** online). More colors may be utilized by varying the $^{12}$C/$^{13}$C ratio during growth (**Supplementary Fig. S14** online), or by monitoring the radial breathing modes of diameter-separated SWNTs[40]. This may lead to many-color SWNT tags for multiplexed protein and biomarker assays[41].

## DISCUSSION

Highly selective and sensitive protein detection has been accomplished via SWNT Raman tags. Their high scattering cross-section, resonant-enhancement, simple spectra, and facile isotopic tuning make SWNTs ideal Raman tags for sensitive detection of proteins and other biomolecules. The development of a branched-PEG coating providing both amine functionality and inert methoxy termini on PEGylated SWNTs enabled production of target-specific, biocompatible nanotubes with minimal non-specific interactions between surfaces, proteins, and SWNTs. Optimization of the SWNT-antibody conjugates required almost two years to develop, as other attempts at functionalization led to unacceptable NSB between proteins and nanotubes, and even small degrees of NSB prevented protein detection at the 1-100 fM level due to false signal.

The coupling of spatially uniform SERS-active gold or silver structures to SWNT Raman tags allowed increased signal-to-noise ratios for protein detection, reducing assay time and improving the limits of detection. The sharp SWNT Raman scattering peaks were not plagued by quenching in proximity to metal surfaces, as is observed with typical fluorescent dyes. This allows for microarrays performed on a variety of metal-coated substrates, utilizing well-established surface chemistry for optimal selectivity and sensitivity of protein detection.

While our current method of SERS (using random metal clusters) allows limits of detection 1000-fold lower than fluorescence-based methodologies, individual SWNT Raman tags likely experience a range of scattering enhancement-factors, depending on the location of nanotubes relative to metal structures, the local field enhancement of SERS hot spots, and the spatial and statistical distribution thereof. Small numbers of the "brightest" SERS hotspots exist and yet may contribute to the majority of the total signal[34]. In SWNT-based protein detection, aggregation of a gold thin-film into nanostructures occurs after the statistical binding of IgG/SWNT tags to immobilized analyte. At relatively high concentrations, SWNTs occupy the small percentage of locations yielding the greatest surface enhancement (i.e. saturation), consequently contributing the majority of the total intensity in a non-linear fashion. At reduced analyte concentrations, the spatial distribution of bound SWNT tags is too few to statistically occupy such hotspots, leading to a proportionally increased contribution of lesser enhanced SWNTs to the average intensity at each concentration. Such dependence yields a dose-response curve demonstrating sensitivity to analyte concentration in a non-linear manner.

While SERS and saturation effects contribute to the non-linearity of dose-response quantification, the observation of sufficiently large surface-enhanced Raman signal change down to 1 fM of model analyte is highly significant, owing to improved signal-to-noise ratios and reduced background intereference. Logistic (4 parameter) curve fitting was employed to aid in accurate quantification of unknown analyte samples, and was proven effective in this regime for quantifying both model analyte, and for a true biomarker of human disease in human serum.

In summary, we used antibody-tagged SWNTs as multi-color Raman labels for detection of proteins in microarrays on various surfaces. Novel functionalization and bioconjugation chemistry of SWNTs, combined with protein immobilization on PEGylated, Raman-enhancing surfaces has allowed application of the sharp, background-free Raman signatures of SWNTs in protein detection, resulting in reproducible sensitivity down to 1 fM (or 0.15 pg/mL) of analyte. In all cases, the SWNT assay was found to be three orders of magnitude greater than standard fluorescence assays. This methodology was then applied to the clinically relevant detection of an autoimmune disease biomarker, anti-Proteinase 3. While this work has focused on antibody-antigen interactions, the application of SWNT Raman tags in specific biomolecule targeting and detection is highly generalizable. Future work and application of ultra-bright, background-free, multi-color SWNT Raman labels is exciting, with a goal of simultaneous detection of multiple analytes in complex fluids, with 1 fM sensitivity in a multiplexed, arrayed fashion.

**METHODS**

**SWNT-antibody conjugates.** SWNTs in this study were either raw HipCO SWNTs (Carbon Nanotechnologies Inc.) or iron ruthenium bimetallic catalyzed[37] chemical vapor deposition (CVD) SWNTs ($^{12}$C- and $^{13}$C-methane[38] as the gas sources respectively). Aqueous SWNT suspensions were prepared by 1 h bath sonication of an aqueous solution of a 1:1 mole ratio mixture of 1,2-distearoyl-*sn*-glycero-3-phosphoethanolamine (DSPE)-PEG$_{5000}$-3PEO ($M_n \sim 8000$) (Polyethylene Oxide) and DSPE-PEG$_{5000}$-NH$_2$ surfactants (NOF corporation), as described previously.[25,26,42]

To conjugate goat anti-mouse IgG (GaM-IgG) or goat anti-human IgG (GaH-IgG, Pierce) to SWNTs, first Traut's Reagent (Pierce) was used to thiolate primary amines on the IgG. Sulfo-SMCC (sulphosuccinimidyl 4-N-maleimidomethyl cyclohexane-1-carboxylate, Pierce) was mixed at a 1:5 molar ratio with the SWNT suspension and incubated at pH 7.4 for 2 h at room temperature. The excess sulfo-SMCC was then removed by centrifugal filtration through a 100 kDa molecular weight cutoff membrane (Millipore). After filtration, a 10-fold molar excess of thiolated GaM-IgG (or GaH-IgG) was added into the SWNT suspension and reacted overnight in PBS at 4 ℃. Excess unconjugated IgGs were removed by filtration. The antibody-conjugated, short SWNTs (50-150 nm long) were characterized by atomic force microscopy (Nanoscope IIIa, Veeco).

**Protein arrays for Raman assay.** Au (5 nm Au/0.2 nm Ti) coated glass slides were washed and oxygen plasma treated (Gabler Labor Instrument) and immersed into a 5 mM cysteamine (Aldrich) ethanol solution overnight. The substrate was then immersed for 2 h in a DMF solution containing 0.1 mM 6arm-PEG-COOH ($M_n$ ~ 10,000, see **Supplementary Information** online for details), 50 mM 1-ethyl-3-(3-dimethylaminopropyl) carbodiimide hydrochloride (EDC, Aldrich) and 50 mM N-hydroxysuccinimide (NHS, Pierce) to form Au surfaces coated with branched PEG (See **Supplementary Fig. S15** online for surface hydrophilicity characterization). Then, 0.5 μL of 1 μM protein (e.g. HSA or PR3) solution was spotted via pipette and incubated on the EDC/NHS activated branched-PEG coating for 2 h in PBS at pH 7.2. The substrates were immersed in 0.1 M tris in PBS solution for 1 h to quench the reaction. The substrates were then immersed into 3% fetal bovine serum (FBS) and 0.5% w/v tween-20 (Aldrich) PBS solution overnight at 4 ℃ for blocking. For multi-color Raman detection, 1 μM purified human IgG and mouse IgG were printed in two sets each of triplicate 400 μm diameter protein spots via solid printing pins (Array-It) with the robotic Bio-Rad VersArray Compact array printer on Superfrost Plus glass slides (Fisher) at 25 ℃ and 65% humidity.

**Protein Raman detection.** For two-layer direct detection of immobilized proteins (**Fig. 2**), 20 nM GaM-IgG/SWNT in PBS (molar extinction coefficient of short SWNTs[43] $\varepsilon_{808nm} \approx 0.0079$ nM$^{-1}$cm$^{-1}$) solution was incubated on the surface for 30 min to allow binding prior to Raman detection. For sandwich detection of aHSA, eight aHSA solutions (10 nM – 1 fM in 3% FBS / PBS) were incubated on the substrate containing an array of HSA spots (above) for 24 hours at 4 ºC. A 3% FBS/PBS solution without aHSA was incubated with an HSA protein spot, and served as an analyte-free control. A hydrophobic PAP marker (Cedarlane Labs) was used to circumscribe the array prior to incubation. Following PBS soaking for 5 min and gently rinsing with PBS and water, 20 nM GaM-IgG/SWNT solution was incubated for 30 mins at room temperature. For aPR3 detection, all aPR3 (Immunovision) solutions were prepared by spiking analyte into 1% normal human serum and 3% FBS in PBS. The total human IgG content in the original aPR3-containing sample, determined by ELISA, was 17 mg/mL. Ten serially-diluted aPR3 serum solutions, as well as an aPR3-negative, dilute normal human serum sample, were incubated on PR3 spots for 24 hour at 4 ºC. 20 nM GaH-IgG/SWNT conjugates were then incubated on the spots, allowing binding before Raman detection.

For SERS, assays on gold substrates were annealed in protective hydrogen gas at 400 °C for 3 minutes to form Au clusters over the substrates, which enhanced the SWNT Raman signals. Alternatively, for assays on glass substrates, as in the multi-color experiments, SERS was afforded by depositing a 5 nm thick Ag film on the substrate by electron beam evaporation. Ag clusters were formed uniformly over the substrate, coating the intact proteins and SWNT-tags. All Raman spectra were collected on a Renishaw micro-Raman instrument (laser excitation wavelength at 785 nm). A 20X objective lens (Leica) was used to focus on the protein spots. A laser spot of ~200 $\mu m^2$ was utilized during the Raman scattering measurements. At each protein spot, at least 20 spectra were recorded at different spatial points to obtain an averaged SWNT Raman scattering intensity, as well as the standard deviation, depicted as error bars. The collection time for each spectrum was 1 second. For Raman imaging of protein array spots (**Fig. 3a**), Raman mapping was carried out using a 20X objective, and the integrated G-band intensity was measured over the whole protein spot (approx. 3 mm$^2$), pixel by pixel (pixel size 50 μm x 50 μm).   Error bars for map analyses were generated as the standard deviation of the means of duplicated assay spots (n=9)

**$^{12}$C and $^{13}$C SWNT direct IgG detection assay.** A solution comprising a mixture of ~5 nM $^{12}$C GaM-IgG/SWNT and ~5 nM $^{13}$C GaH-IgG/SWNT conjugates in PBS was incubated over robotically-printed polyclonal mouse and human IgG spots on Superfrost Plus glass slides (Fisher) for 30 min at room temperature. Following Ag deposition, mapping of the SWNT G band intensity over the 2 mm x 3 mm area was performed with 30 μm steps in x and y. The MIgG and HIgG mapping image was generated by integration of the respective G-bands of $^{12}$C and $^{13}$C SWNTs above background, and plotted as false-colored maps (**Fig. 6c**).

**Standard fluorescence-based protein microarray assay.** For fluorescence-based[16] aHSA and aPR3 detection, arrays of six duplicate protein spots were printed robotically. HSA was printed at 1 μM in PBS on Super-Epoxide 2 (Array-It) slides, and PR3 was printed at 0.2 mg/mL in PBS onto nitrocellulose-coated FAST slides (Whatman). Arrays were blocked for 1 h at room temperature in PBS with 0.05% Tween-20 and 3% FBS, and then incubated for 3 - 24 hours at 4 °C with the indicated dilutions of aHSA (in 3% FBS/PBS) or aPR3 (in 1% human serum and 3% FBS/PBS), in addition to unspiked serum controls. Following washing, arrays were incubated with 400 μL of a 10 pM solution of GaM-IgG/Cy3 or GaH-IgG/cy3 in[16] PBS for 1 h at room temperature in the dark. The slides were scanned using a GenePix 4000B Scanner (Axon Labs) with a 17 mW, 532 nm excitation laser and photo-multiplier tube gain set to 400 for nitrocellulose-coated slides (Whatman) and 800 for Super-Epoxide II Slides (Array-It). Array spot features were automatically selected at 10 μm pixel resolution by GenePix Pro 6.0 software (Axon Labs). Reported fluorescence values were calculated as the local background corrected (See **Supplementary Information** online) mean of all pixels identified as array features for each analyte dilution point. Error bars represent the standard deviation between duplicate spots from the same assay substrate.

*Note: Supplementary information is available on the Nature Biotechnology website.*

**ACKNOWLEDGEMENTS**
This work was supported by the NIH/NCI funded Center for Cancer Nanotechnology Excellence Focused on Therapeutic Response U54 CA119367 at Stanford University. The authors would like to thank Nozomi Nakayama-Ratchford and Dr. Sarunya Bangsaruntip for their assistance in developing carbon nanotube-protein conjugates.

**COMPETING INTERESTS STATEMENT**
The authors declare that they have no competing financial interests.

**Figure Captions:**

**Figure 1** Carbon nanotubes as Raman labels for protein microarray detection. (a) A schematic of surface chemistry used for immobilization of proteins on Au coated glass slides for Raman detection of analytes by single-walled carbon nanotube (SWNT) Raman-tags. A self-assembled monolayer of cysteamine on Au was covalently linked to 6-arm branched poly(ethylene glycol)-carboxylate (6arm-PEG-COOH, inset) to provide a layer with high resistance to protein non-specific binding, and terminal carboxylate groups for immobilizing proteins. (b) Sandwich assay scheme. Immobilized proteins in a surface spot were used to capture an analyte (antibody) from a serum sample. Detection of the analyte by Raman scattering measurement was carried out following incubation of SWNTs conjugated to goat anti-mouse antibody (GaM-IgG/SWNTs), specific to the captured analyte. SWNTs were functionalized by (1,2-distearoyl-sn-glycero-3-phosphoethanolamine-poly(ethylene oxide) (DSPE-3PEO) and 1,2-distearoyl-sn-glycero-3-phosphoethanolamine (DSPE-PEG$_{5000}$-NH2) (left panel). (c) SWNT Raman spectra in the G mode and radial breathing mode (RBM, inset) regions before and after SERS enhancement.

**Figure 2** Highly-selective recognition of surface-bound proteins by SWNT-antibody conjugates. (a) A schematic illustration of 2-layer, direct assay of mouse IgGs, which are immobilized with a complex set of protein analytes including BSA, Human IgGs and Avidin, arrayed on a Au/glass substrate by detecting GaM-IgG/SWNT Raman intensities. (b) G-mode intensities of GaM-IgG/SWNT conjugates bound to various protein spots on the substrate. Specific target proteins that were detected included polyclonal MIgG (Sigma), mouse anti-PSA (anti prostate specific antigen, Fitzgerald), mouse anti-TNF (Tumor Necrosis Factor, Pepro Tech. Inc.), mouse aHSA (Medix MAB Inc.), mouse aIL1 (anti interleukin-1, Fitzgerald) and mouse aHCG (anti human chorionic gonadotropin, Biospacific). Negative control proteins included human IgG (HIgG, Sigma), Streptavidin (SA, Pierce), PSA (Fitzgerald), BSA (bovine serum albumin, Pierce), HSA (Sigma), HCG (Biospacific), TNF (Pepro Tech. Inc.), AV (Avidin, Sigma). PBS (Fisher) was also spotted on the surface as a control.

**Figure 3** Femtomolar protein detection using SWNT Raman labels, in comparison with fluorescence-based protein microarray detection. (a) Raman mapping images showing integrated SWNT G-band peaks for model protein sandwich assays of anti-human serum albumin (aHSA) in fetal bovine serum (FBS) ranging in concentration from 100 pM to 1 fM and an FBS control. Scale bar is 300 μm. (b) A log-log plot of the G-band Raman intensity vs. aHSA concentration from two separate trials (blue and green) performed using different substrates and different GaM-IgG/SWNT conjugate batches on different days. A sigmoidal dependence was observed (and fit by four-parameter logistic function, red curve) suggesting that protein quantification by Raman scattering of SWNT-tags is limited by surface-receptor saturation and steric hindrance at high concentrations, and by residual NSB at the lower detection limit. (c) Fluorescence-based microarray detection of aHSA. The image shows GaM-IgG/Cy3 fluorescence levels of HSA array spots exposed to aHSA ranging in concentration from 1 nM – 1 fM (a sigmoidal dependence, modelled by four parameter logistic fit, blue curve, was observed). The fluorescence of aHSA bound to HSA on the substrate was also recorded without exposure to GaM-IgG/Cy3 as a measurement of autofluorescence and background noise. (d) Mean fluorescence vs. analyte concentration from 1 nM to 1 fM. FBS-only and fluorophore-free controls are also included in the plot. Fluorescence observed for HSA array spots exposed to aHSA at concentrations below 1 pM was not distinguishable from background fluorescence (without exposure to the Cy3 fluorophore), indicating that sensitivity was limited by poor signal-to-noise.

**Figure 4** Calibration curve of mouse anti-human serum albumin measured in microarray format from nine duplicate protein spots at each analyte concentration by SWNT Raman tags. A limit of detection at 1 fM is observed with a dynamic range greater than 6 orders of magnitude. A sigmoidal curve was obtained by four-parameter logistic function fitting of the data, red curve. Blind unknown analyte samples were prepared at 5 pM, 200 fM, and 5 fM, and in comparison with the calibration curve, they were accurately quantified as 2 pM, 110 fM, and 4 fM respectively with this methodology.

**Figure 5** Raman vs. fluorescence-based protein microarray detection of aPR3, a biomarker for Wegener's granulomatosis (WG), in human serum. (a) Schematic illustration of specific detection of aPR3 spiked into dilute human serum. aPR3 is captured by PR3 antigen on a PEGylated, gold-coated substrate, and detected by anti-human IgG conjugated SWNT Raman tags. (b) SWNT G-mode intensity as a function of aPR3 analyte concentration, captured from dilute human serum. The data shows two separate sensing trials, using different assay substrates and different batches of GaH-IgG/SWNTs. Deviation in signal intensity is systematic, and likely a product of deviation in gold-thickness on the assay substrate and slight variation in the loading efficiency of anti-human IgG to the nanotube tags from batch to batch. Our assay design, like most bioassays, requires the use of simultaneous calibration curve measurements on the same chip. Accurate determination of unknown concentrations is possible because all measurements, standards and unknowns, are performed with the same reagents on the same assay substrate (See **Figure 4** for such an example). A sigmoidal dependence was observed (and fit by four-parameter logistic function, red curve) suggesting that protein quantification by Raman scattering of SWNT-tags is limited by surface-receptor saturation and steric hindrance at high concentrations, and by residual NSB at the lower detection limit. (c) Mean fluorescence as a function of aPR3 concentration captured from dilute human serum in a microarray experiment parallel to that described in (b) (a sigmoidal dependence, modelled by four parameter logistic fit, blue curve, was observed).

**Figure 6** Multi-color SWNT Raman labels for multiplexed protein detection. (a) Schematic illustration of two-layer, direct, microarray-format protein detection with distinct, Raman labels based upon pure $^{12}$C and $^{13}$C SWNT tags. $^{12}$C and $^{13}$C SWNTs were conjugated to GaM and GaH-IgGs, respectively, providing specific binding to complimentary IgGs of mouse or human origin, even during mixed incubation with analyte (as shown). (b) G-mode Raman scattering spectra of $^{12}$C (red) and $^{13}$C (green) SWNT Raman-tags are easily resolvable, have nearly identical scattering intensities, and are excited simultaneously with a 785 nm laser. This allows rapid, multiplexed protein detection. c) Raman scattering map of integrated $^{12}$C (red) and $^{13}$C (green) SWNT G-

mode scattering above baseline, demonstrating easily resolved, multiplexed IgG detection based upon multi-color SWNT Raman labels.

**Figure 1**

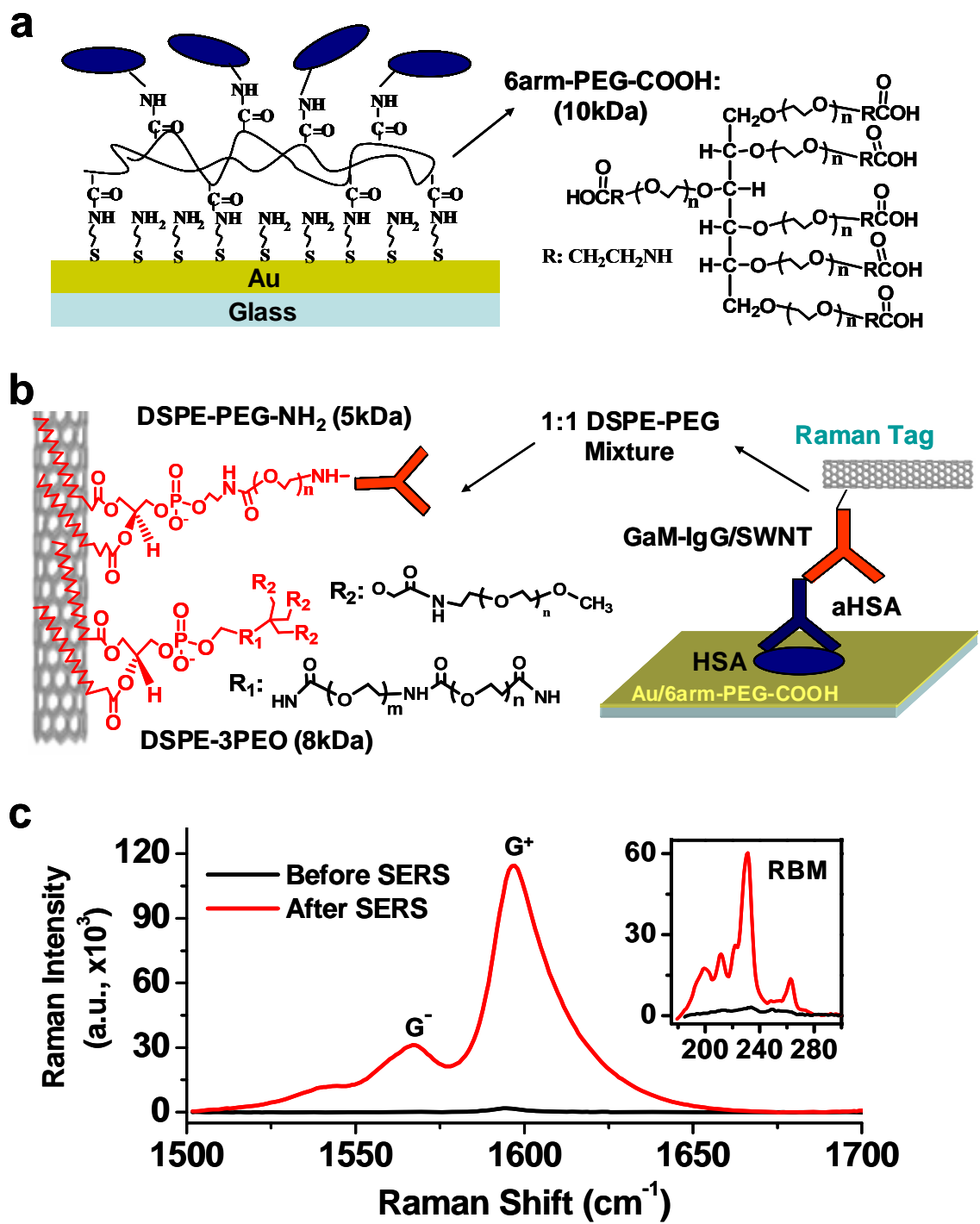

**a**

6arm-PEG-COOH:
(10kDa)

R: CH₂CH₂NH

**b**

DSPE-PEG-NH₂ (5kDa)

1:1 DSPE-PEG
Mixture

Raman Tag

GaM-IgG/SWNT

aHSA

HSA

Au/6arm-PEG-COOH

$R_2$: 

$R_1$: 

DSPE-3PEO (8kDa)

**c**

**Figure 2**

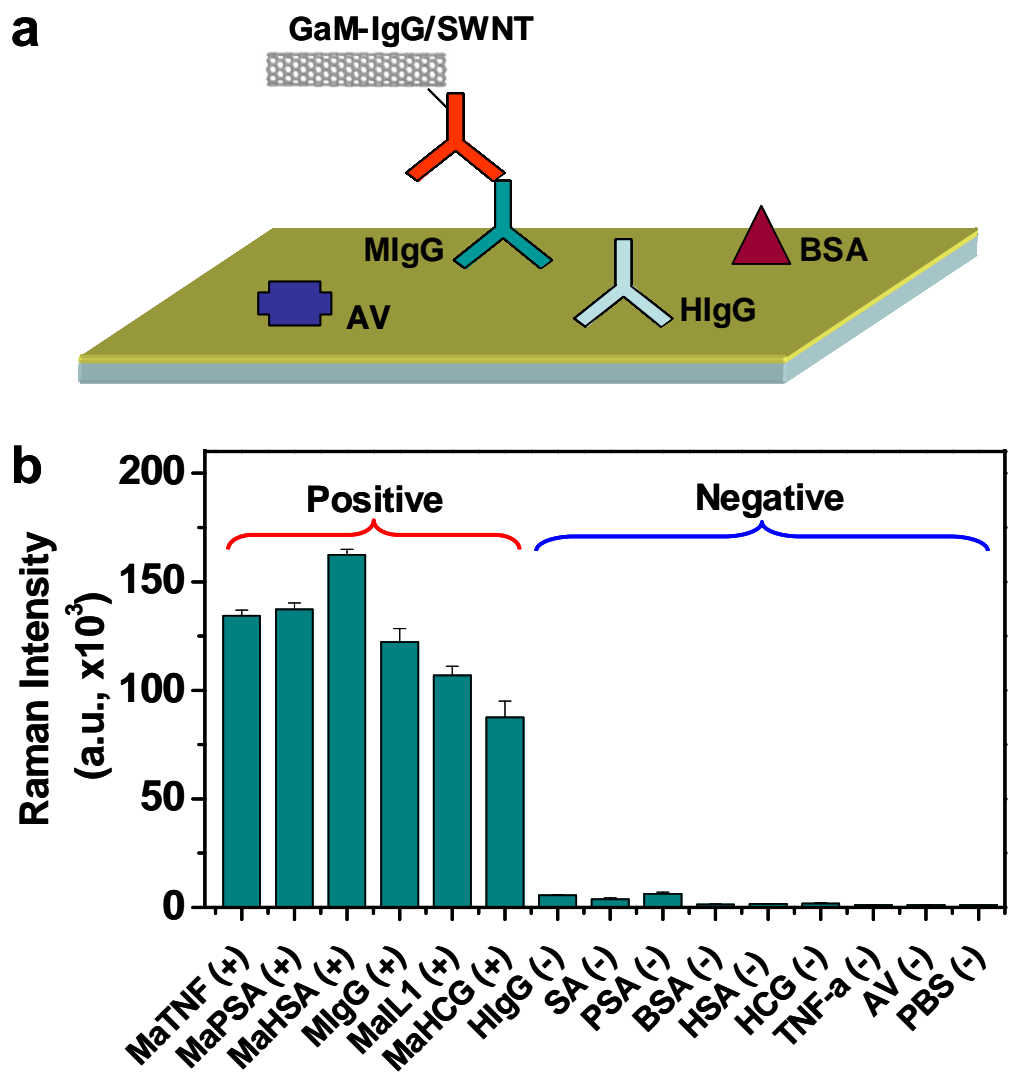

**Figure 3**

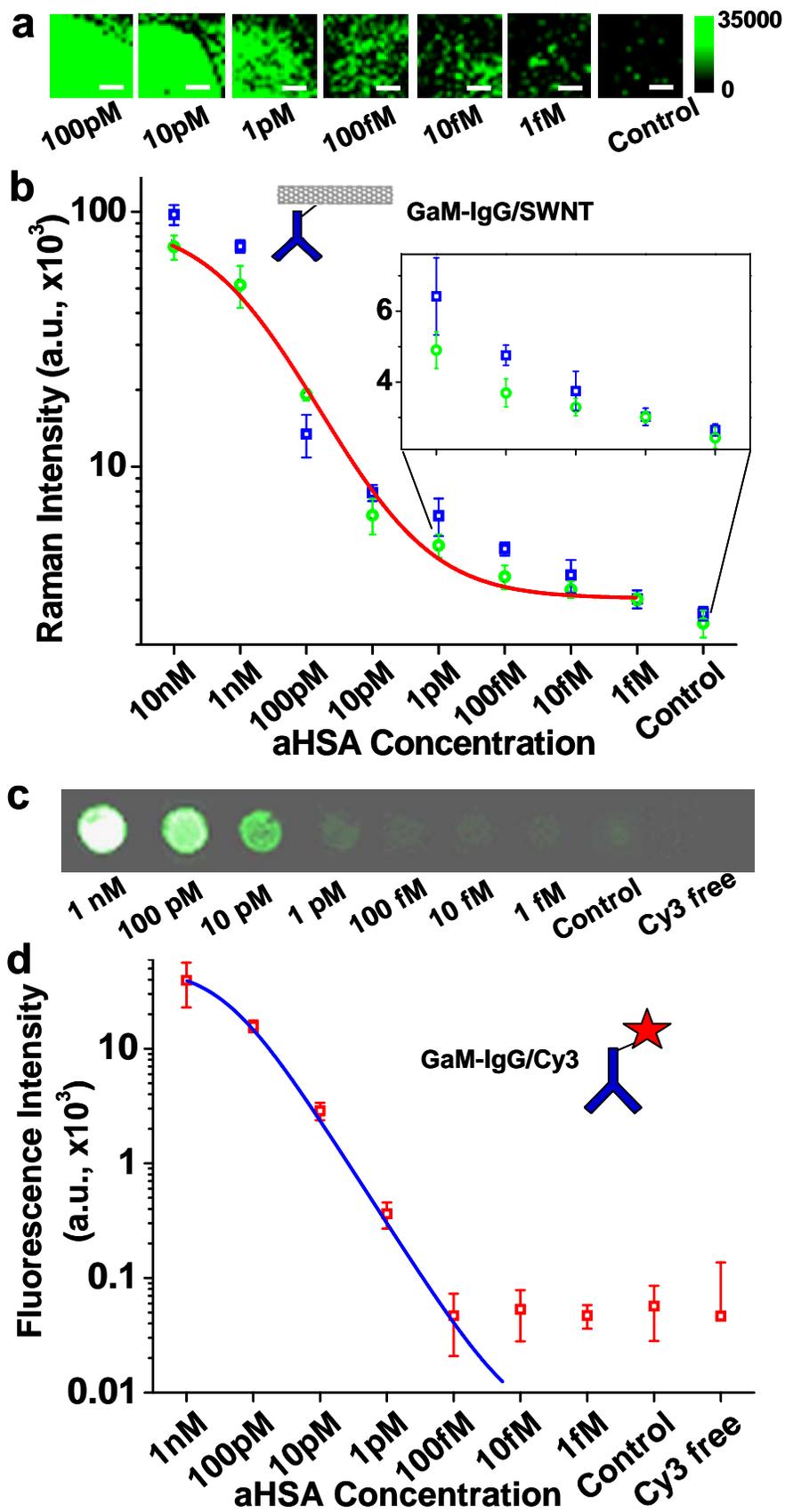

**Figure 4**

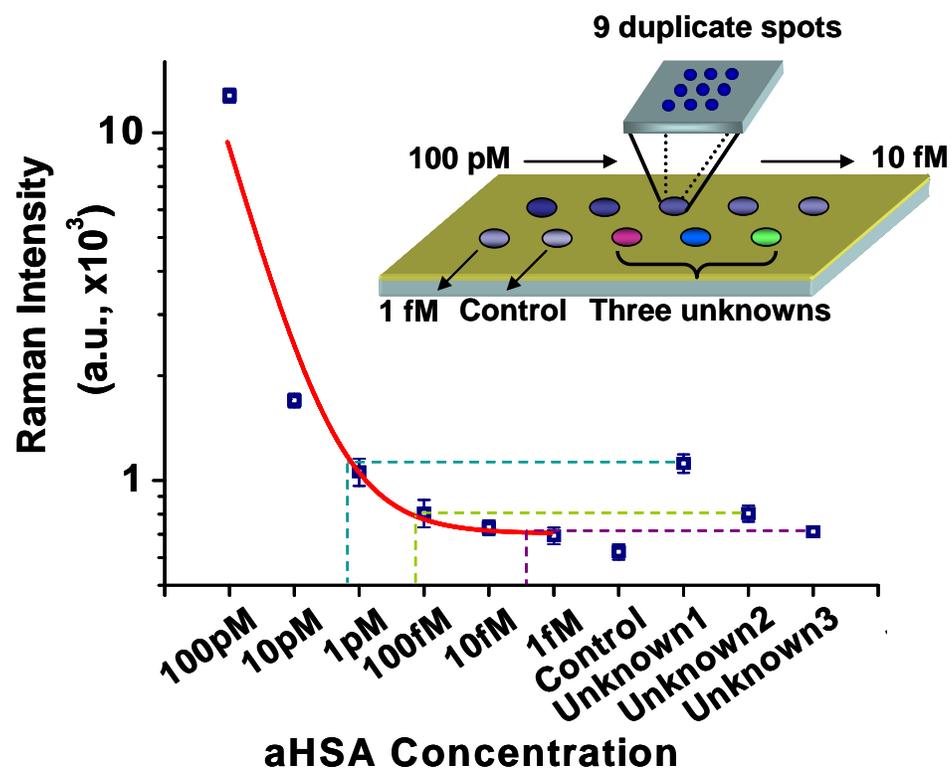

Figure 5

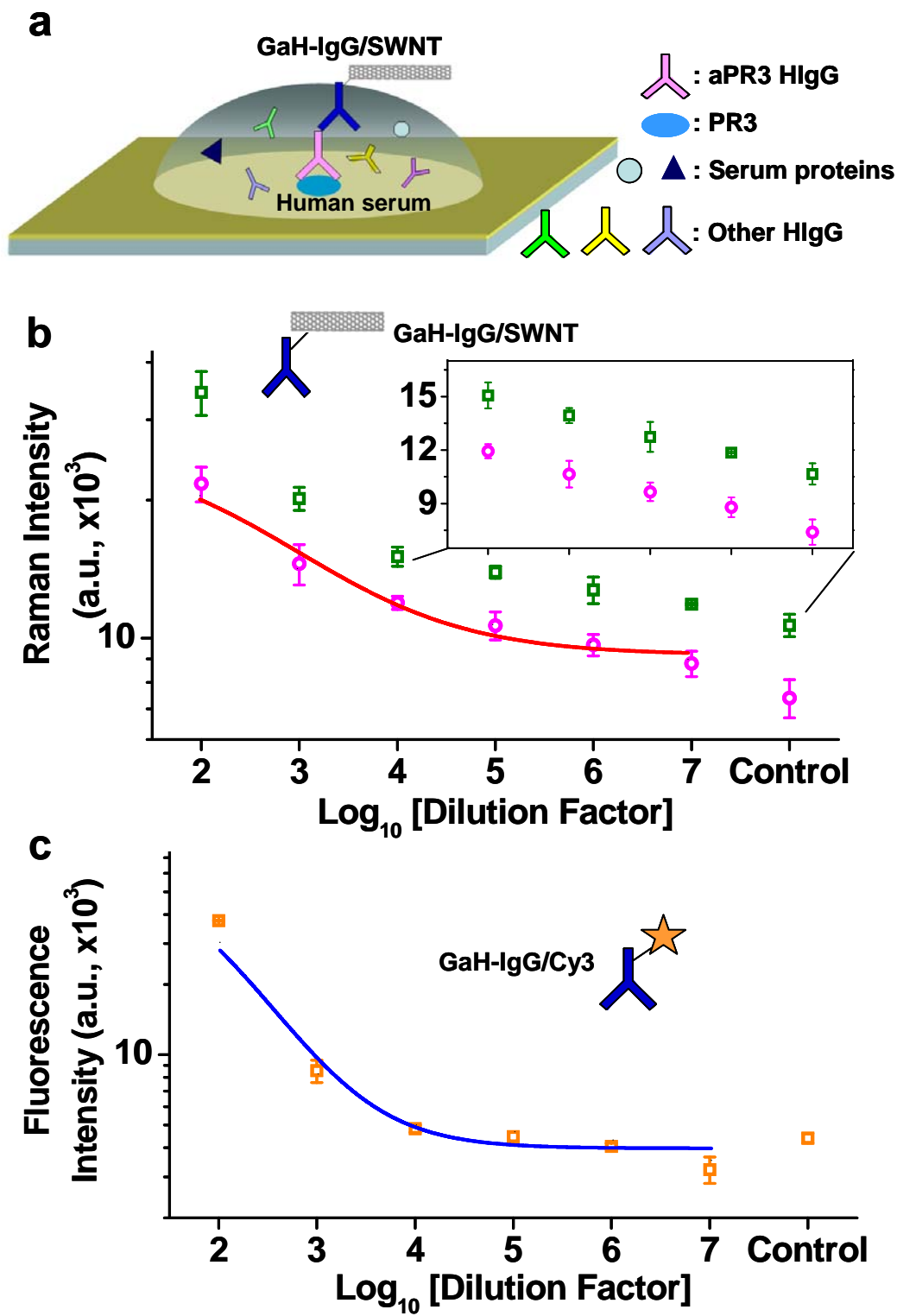

Figure 6

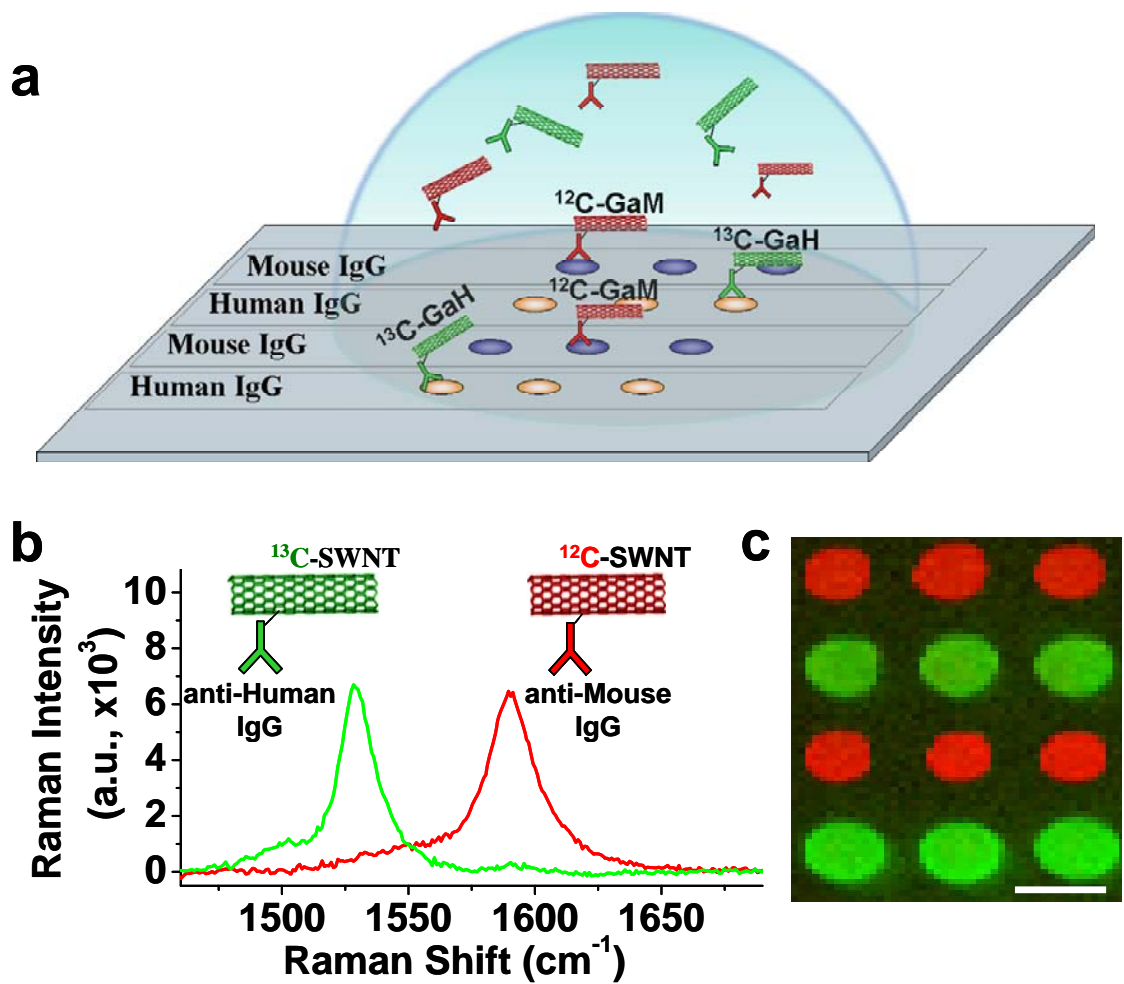